\documentclass[aps,nofootinbib,twocolumn,groupedaddress]{revtex4}
\usepackage{graphicx}
\usepackage{color}
\begin{document}

\title{Branching ratio approximation for the self-exciting Hawkes process}

\author{Stephen J. Hardiman}
\author{Jean-Philippe Bouchaud}
\affiliation{
 Capital Fund Management, 23 rue de l'Universit\'e, 75007 Paris, France\\
}

\date{\today}

\begin{abstract}
We introduce a model-independent approximation for the branching ratio of Hawkes self-exciting point processes. Our estimator requires knowing only the mean and variance of the event count in a sufficiently large time window, statistics that are readily obtained from empirical data. The method we propose greatly simplifies the estimation of the Hawkes branching ratio, recently proposed as a proxy for market endogeneity and formerly estimated using numerical likelihood maximisation. We employ our new method to support recent theoretical and experimental results indicating that the best fitting Hawkes model to describe S\&P futures price changes is in fact critical (now and in the recent past) in light of the long memory of financial market activity.
\end{abstract}

\maketitle

\section{The self-exciting Hawkes Process}

The Hawkes model \cite{hawkes,hawkes2} is a simple and powerful framework for simulating or modelling the arrival of events which cluster in time (e.g. earthquake shocks and aftershocks, neural spike trains and transactions on financial markets). In one dimension, the model is a counting process $N(t)$ with an intensity $\lambda(t)$ (the expected number of events per unit time) given by a constant term $\mu$ and a `self-exciting' term which is a function of the event history.

\begin{equation}
\lambda(t) = \mu + \int_{-\infty}^t \phi(t-s) {\rm d}N(s) \label{eq:hawkes}
\end{equation}

This self-exciting term gives rise to event clustering through an endogenous feedback: past events contribute to the rate of future events. $\phi(\tau) \geq 0$ is the ``influence kernel'' which decides the weight to attribute to events which occurred at a lag $\tau$ in the past. The base intensity $\mu$ and the kernel shape $\phi(t)$ are parameters to be varied. A popular choice for the kernel is the exponential function $\phi(\tau) = \alpha e^{-\beta \tau}$ \cite{filimonov,ozaki} but in general the kernel to be used should depend on the application or the dynamics of the data to be modelled. Note that for $\phi(t) = 0$ the model reduces to a Poisson process with constant intensity $\mu$.

By taking the expectation of both sides of Eq. (\ref{eq:hawkes}) and assuming stationarity (i.e. a finite average event rate $E[\lambda(t)] = \Lambda$), we can express the average event rate of the process as $\Lambda = \mu/(1-n) \geq \mu$ where $n = \int{\phi(\tau)}{\rm d }\tau$. One can create a direct mapping between the Hawkes process and the well known branching process \cite{harris} in which exogenous ``mother'' events occur with an intensity $\mu$ and may give rise to $x$ additional endogenous ``daughter'' events, where $x$ is drawn from a Poisson distribution with mean $n$. These in turn may themselves give birth to more ``daughter'' events, etc.

The value $n$, which corresponds with the integral of the Hawkes kernel is the branching ratio, determines the behaviour of the model. If $n > 1$, meaning that each event typically triggers at least one extra event, then the process is non-stationary and may explode in finite time \cite{commodityreflexivity}. However, for $n < 1$, the process is stationary and has proven useful in modelling the clustered arrival of events in a wide variety of applications including neurobiology \cite{neurobiology}, social dynamics \cite{social,crime} and geophysics \cite{earthquakes1,earthquakes2}. The Hawkes model has also seen many recent applications to finance \cite{bauwens,hawkesvolatility,bormetti}, especially as a means of modelling the very high-frequency events affecting the limit-order book of financial exchanges \cite{bacry,hawkesmicrostructure,bacrypricetrades,toke,dafonseca}.

One novel application of the Hawkes framework to finance is as a means of measuring market endogeneity or `reflexivity' in financial markets \cite{filimonov,commodityreflexivity}. In \cite{filimonov}, the authors consider mid-price changes in the E-mini S\&P Futures contract between 1998 and 2010 and observe that the branching ratio $n$ of the best-fitting exponential kernel model has been increasing steadily over this period, from $n \approx 0.25$ in 1998
to $n \approx 0.65$ in 2010 (see our Fig. 6 below). They argue that this observation implies that the market has become more reflexive in recent years with the rise of high frequency and algorithmic trading and is therefore more prone to market instability and so-called ``flash crashes''.

In \cite{criticalreflexivity}, however, we have argued that due to the presence of long-range dependence in the event rate of mid-price changes (detectable in both 1998 and 2011) as one extends the window of observation, the best fitting stationary Hawkes model must in fact be critical, i.e. have a branching ratio $n=1$. This is backed up by theoretical arguments and empirical measurements on market data.

Let us however insist that this conclusion only holds if one believes that Hawkes processes provide an exact representation of the reality of markets. It is very plausible that the dynamics of markets is more complicated (and involves, for example, non-linearities absent from the Hawkes process defined by Eq. (\ref{eq:hawkes})), but that the best way to represent this dynamics within the framework of Hawkes processes is to choose $n=1$ with a long-ranged influence kernel.

In this article we introduce a simple approximation for the branching ratio of the Hawkes process which allows us to faithfully reproduce the results of \cite{filimonov} which proposes the statistic as a measure of market instability and as a crash prediction metric. The interest of our approximation lies in its great simplicity: one need only estimate the mean and variance of the event count in a sufficiently large time window. The approximation also avoids a number of pitfalls \cite{apparentcriticality} inherent to the significantly more complex approach \cite{ozaki} employed in \cite{filimonov}.

The estimator accepts one parameter, a time window size, $W$, during which we measure the mean and variance of the event count. We note that when we employ our estimator to mid-price changes in the S\&P electronic futures market with a fixed window size $W$ then the branching ratio estimate obtained increases over time as reported in \cite{filimonov}. If, however, we allow the window size to scale appropriately (halving in size every 18 months) to adapt to the decreasing latency of interactions on the market we recover a constant branching ratio estimate as proposed in \cite{criticalreflexivity}. This result reiterates the need for a scale-invariant, or at least scale-sensitive means of measuring the `reflexivity' of financial market events.

\section{Maximum likelihood estimation}

Given observed events (e.g. mid-price changes) at times $t_1,t_2, \ldots, t_n$ in an interval $[0,T]$ one can fit the Hawkes model by maximising the log-likelihood \cite{rubin,ozaki} over the set of parameters $\theta$.

\begin{equation}
\log L(t_1,\ldots,t_n| \theta) = -\int_0^T \lambda(t| \theta) {\rm d}t + \int_0^T \log \lambda(t|\theta) {\rm d}N(t) \label{eq:loglikelihood}
\end{equation}

In the case of the exponential kernel $\theta = \{\mu,\alpha,\beta\}$. In practice, the model parameters $\{\hat{\mu},\hat{\alpha},\hat{\beta}\}$ which maximise this log-likelihood are obtained with numerical techniques due to the lack of a closed form solution \footnote{Note that the method above is not the only method proposed to fit the parameters of the Hawkes process for financial applications, indeed a recent publication \cite{dafonseca} proposes a fast albeit still parametric method for fitting the multivariate exponential Hawkes process.}. The branching ratio estimate is then $\hat{n} = \hat{\alpha}/\hat{\beta}$.

However, there are a number of pitfalls to using this procedure as a means of estimating the Hawkes branching ratio $n = \int \phi(\tau) {\rm d}\tau$ \cite{apparentcriticality}. Arguably the most important of which is that any estimate of $n$ made in this manner will be heavily dependent on the choice of kernel model (e.g. exponential, power-law, etc.) It may be that the chosen model cannot satisfactorily describe the observed events, hence the meaning of the branching ratio extracted from the maximum likelihood method is questionable.

Care must also be taken when employing this method in the presence of imperfect event data as illustrated in \cite{apparentcriticality}. In one of their figures, the authors present a (negative) log-likelihood surface which features two minima (one local, and one global). The global log-likelihood minimum does in fact little more than describe packet clustering inside the millisecond which arises from the manner in which events arriving at the exchange at different times are bundled and recorded with the same timestamp. Subsequent randomisation of the timestamps inside a short time interval (in this case, one millisecond) creates a spurious high frequency correlation, that makes the global minimum irrelevant. The local log-likelihood minimum, which is in fact a better fit to the `true' lower frequency dynamics, does a poor job at explaining the spurious high frequency clustering and is punished with a lower log-likelihood. Indeed, when the authors of \cite{apparentcriticality} choose to fit this local minimum they corroborate results presented in \cite{criticalreflexivity}.

We believe it is therefore essential to have additional checks (such as non-parametric methods \cite{bacry}) at one's disposal to support results obtained from likelihood maximisation. To address the pitfalls in branching ratio estimation that arise from the model choice we propose a simple model-independent tool for branching ratio approximation, in the next section, which accurately reproduces previous results of \cite{filimonov} and also indicates the criticality of the relevant Hawkes process which describes the market.

\section{A mean-variance estimator for the branching ratio $n$}

We begin with a general expression relating the Fourier transform of the kernel function to the Fourier transform of the auto-covariance $\nu(\tau) = E[\frac{{\rm d}N(t){\rm d}N(t+\tau)}{{\rm d}t^2}] - \Lambda^2$ of the event rate. (see \cite{hawkes, bacry} for a derivation).

\begin{equation}
\hat{\nu}(\omega) = \frac{\Lambda}{\left|1-\hat{\phi}(\omega)\right|^2}
\end{equation}
Setting $\omega = 0$ we obtain a relation between the branching ratio, the average event rate and the integral of the auto-covariance (in the stationary case $n \leq 1$)
\begin{equation}\label{basic}
\int_{-\infty}^\infty \nu(t) {\rm d}t = \frac{\Lambda}{\left|1-\int_{0}^\infty \phi(t) {\rm d}t\right|^2} \equiv \frac{\Lambda}{(1-n)^2}
\end{equation}

Therefore, to deduce the branching ratio of the stationary Hawkes process, we need only {find the value of} $\Lambda$ and $\int_{-\infty}^\infty \nu(t) {\rm d}t$. Estimating $\Lambda$ is trivial, it is given by the
empirical average number of events per unit time. To estimate $\int_{-\infty}^\infty \nu(t) {\rm d}t$, we consider the variance {of the total event count $N_W$} in a window of length $W$. Theoretically, 
this is given by:

\begin{eqnarray}
\sigma^2_W &=& \mathrm{E} \left[\int_{t=0}^W \frac{{\rm d}N(t)}{{\rm d}t} {\rm d}t \int_{t'=0}^W \frac{{\rm d}N(t')}{{\rm d}t'} {\rm d}t'\right] - (\Lambda W)^2 \nonumber \\
&=& \int_{t=0}^W \int_{t'=0}^{W} \nu(t-t') {\rm d}t \, {\rm d}t' \nonumber \\
&=& \int_{t=0}^W \int_{\tau=-t}^{W-t} \nu(\tau) {\rm d}t \, {\rm d}\tau \nonumber \\
&=& \int_{\tau=-W}^W \nu(\tau) \left(W-\left|\tau\right|\right) {\rm d}\tau \nonumber \\
&\leq& W \int_{-\infty}^\infty \nu(\tau) d\tau \label{eq:bounded}
\end{eqnarray}

We then note that provided:
\begin{enumerate}
\item $\nu(t) \to 0$ for $|t| > R$
\item $W \gg R$ such that $(W-|t|)/W \approx 1$ for all $|t| < R$
\end{enumerate}
then
\begin{equation}
\sigma^2_W \approx W \int_{\tau=-\infty}^\infty \nu(\tau) {\rm d}\tau \label{eq:covariance_integral}
\end{equation}
and we find, using Eq. (\ref{basic}):
\begin{eqnarray}
n \approx& 1 - \sqrt{\frac{{\mu_W}}{{\sigma^2_W}}} := \tilde{n} \label{eq:simple_estimator}
\end{eqnarray}
{where $\mu_W = \Lambda W$ is the average number of events falling in a window of size $W$. The above expression} has a very intuitive interpretation. When the variance is equal to the event rate, the process is obeying Poisson statistics and $n = 0$. Any increase in the variance above the event rate indicates 
some positive correlations and, within a Hawkes framework, a positive branching ratio.

Note from Eq. (\ref{eq:bounded}) that $\sigma^2_W / W$ is a biased estimator for $\int_{-\infty}^\infty \nu(\tau)$ and for a general $W$ will fall short of its theoretical value. This means that Eq. (\ref{eq:simple_estimator}) will generally under-estimate the branching ratio and only become exact in the limit $W \to \infty$.

{In practice, to estimate the branching ratio of an empirical sample of total length $T = mW$ we substitute the mean and variance term 
in Eq. (\ref{eq:simple_estimator}) by their sample estimates: 
\begin{eqnarray}
\tilde{\mu}_W = W \frac{N_T}{T} \equiv \frac{1}{m} \sum_{i=1}^{m} N_W(i) \\
\tilde{\sigma}^2_W = \frac{1}{m-1} \sum_{i=1}^{m} \left(N_W(i) -\tilde{\mu}_W\right)^2 \label{eq:variance}
\end{eqnarray}
An estimate obtained in this fashion is asymptotically convergent for large $T \gg W$, i.e. $m \gg 1$. For a fixed window size $W$ we can always ensure that our estimate of the variance of $N_W$ is within a desired maximum 
error with a desired minimum probability by increasing $T$ and therefore the number of observations $m$ of $N_W$. Furthermore, we can make Eq. (\ref{eq:covariance_integral}) exact by allowing $W$ to be arbitrarily large.}

{Note, however, that for any finite $m$, the average of $\tilde{n}$ over all possible realisations of the process is in fact $-\infty$! This is because a value $\tilde{\sigma}^2_W = 0$ is always possible with a non zero probability. For this reason we choose to present the median of our estimates.}

\section{Numerical simulations \& Implementation notes}
\begin{figure}[h]
\includegraphics[width=1\columnwidth]{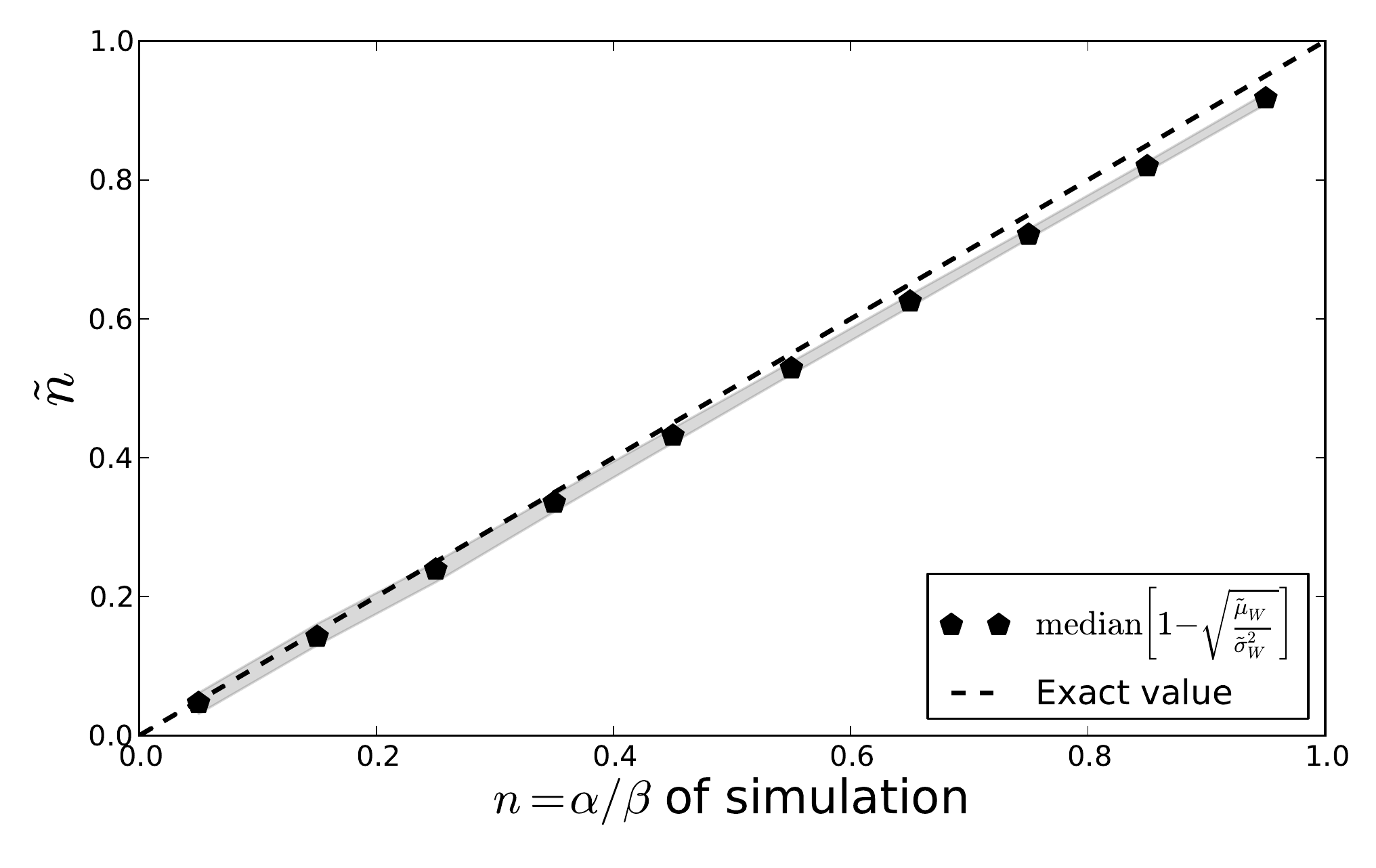}
\caption{\label{fig:varying_n} Applying the mean-variance branching ratio approximation method to a simulated Hawkes process with exponential kernel form and scale parameter $\beta = 1.0$. $\alpha$ is varied to decide the branching ratio, and $\mu$ is varied to keep the average event rate fixed at $\Lambda = 1.0$. The process is simulated for $T = 100000s$ and the branching ratio estimate is made using Equation \ref{eq:simple_estimator} with a window size $W = 20$. The mean and [5\%,95\%] confidence bands are calculated over an ensemble of 100 process for each parameter set.}
\end{figure}

{
In Fig. \ref{fig:varying_n}, we test the estimation procedure described in the previous section on a number of simulated exponential Hawkes processes with a variety of branching ratios. To do this we fix $\beta = 1.0$ but vary $\alpha = n$ in the range $0 \leq \alpha \leq 0.95$. We choose a base intensity $\mu = 1-n$ such that each process has the same average event rate $\Lambda = 1$. We simulate the process for a time $T = 10^5$ using simulation Algorithm 1 described in \cite{moller2005perfect}. This procedure is prone to edge effects so we disregard the initial period of size $10^4$ to ensure the process is close to stationarity in the period studied.
}

{
We estimate the branching ratio with a window size $W = 20$ significantly larger than the time-scale of correlation of our process $\beta^{-1} = 1$. However, the window size chosen is also sufficiently small that we have a significant number of independent observations $m = (0.9 \times 10^5) / 20$ with which to make a reliable estimate of the variance of $N_W$. We see very good agreement between our branching ratio estimates and the input branching ratio of the simulation, see Fig. \ref{fig:varying_n}.
}

Note however, that our approximation systematically underestimates the branching ratio since our finite window size does not completely cover the region of support of the autocorrelation function. This is particularly visible in Fig. \ref{fig:varying_n} for large $n$. We now investigate the effect of window size on the estimate obtained.

\begin{figure}[h]
\includegraphics[width=1\columnwidth]{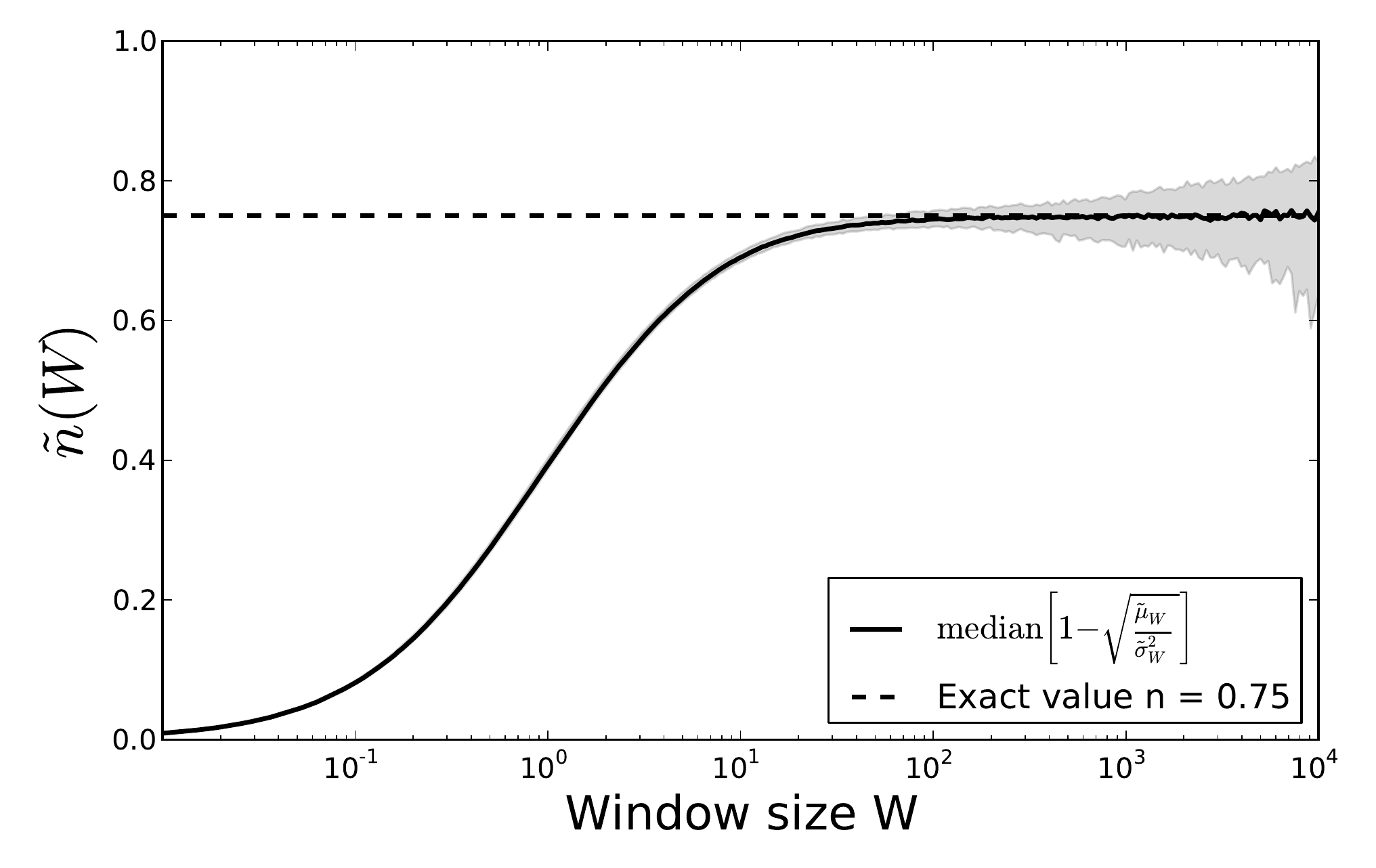}
\caption{\label{fig:varying_W_simulation} Applying the mean-variance branching ratio approximation method to simulated data. Shaded area represents the [5\%,95\%] confidence interval. We note that the branching ratio estimate converges on the expected value of $0.75$ as the window size increases. For very large W, we lack a sufficient number of event count observations to estimate the variance of $N_W$ with precision and the confidence interval grows considerably.}
\end{figure}

{
\subsection{Choice of window size W}
For Eq. (\ref{eq:covariance_integral}) to be accurate, we must choose a large $W$. However for a finite sample size, large $W$ implies small $m = (T/W)$ and therefore less statistical power with which to estimate the variance of the event count $N_W$. This compromise is illustrated in Fig. \ref{fig:varying_W_simulation} for which we simulate an ensemble of exponential Hawkes processes with parameters $\alpha=0.75$, $\beta=1.0$ and $\mu = 0.25$. We note that for increasing window size $W$, the confidence bands of our estimate converge on the expected value of $n = 0.75$.
}

However when we make the window size too large, we suffer significant error coming from the estimation of the variance. For the practical implementation of this procedure to empirical data we recommend a choice of window size sufficiently large to capture the bulk of autocorrelation present in the event rate, but sufficiently small that one can expect to obtain reliable estimates of the variance of the event count in that window.

One can approximate upper and lower confidence intervals on the branching ratio estimate from a single realisation of a time series by bootstrap re-sampling. Indeed the simple variance estimator of Eq. (\ref{eq:variance}) is not optimal and can be improved with the use of over-lapping windows or, for example, a Monte Carlo sampling scheme which selects random windows with replacement.

{
\subsection{Power-law kernel}
Finally we test the estimation procedure on a (near) critical power-law Hawkes process, a type suggested in some recent publications as a good fit to financial event data \cite{bacry,criticalreflexivity}.
Specifically, we consider a kernel with an Omori law form:
\[
\phi(\tau) = \frac{n \epsilon \tau_0^{\epsilon}}{\left(\tau_0 + \tau\right)^{\left(1+\epsilon\right)}}
\]
}

{
In the critical case of $n=1$ with $0 < \epsilon < 0.5$ this process will exhibit long-range dependence, with an autocorrelation function $\nu(\tau)$ decaying asymptotically as a power law: $\nu(\tau) \sim \tau^{-(1-2\epsilon)}$ \cite{bremaud,criticalreflexivity}. The integral of the autocorrelation function is therefore divergent for large $\tau$'s and the variance of the event count in a window of size $W$ grows with as $\sigma^2_W \sim W^{1 + 2\epsilon}$. The $\sqrt{\frac{\mu_W}{\sigma^2_W}}$ term in Eq. \ref{eq:simple_estimator} will in this case not converge to a finite constant for large $W$ but rather shrink to zero, leading to $1-\tilde{n} \sim W^{-\epsilon}$.
}

\begin{figure}[h]
\includegraphics[width=1\columnwidth]{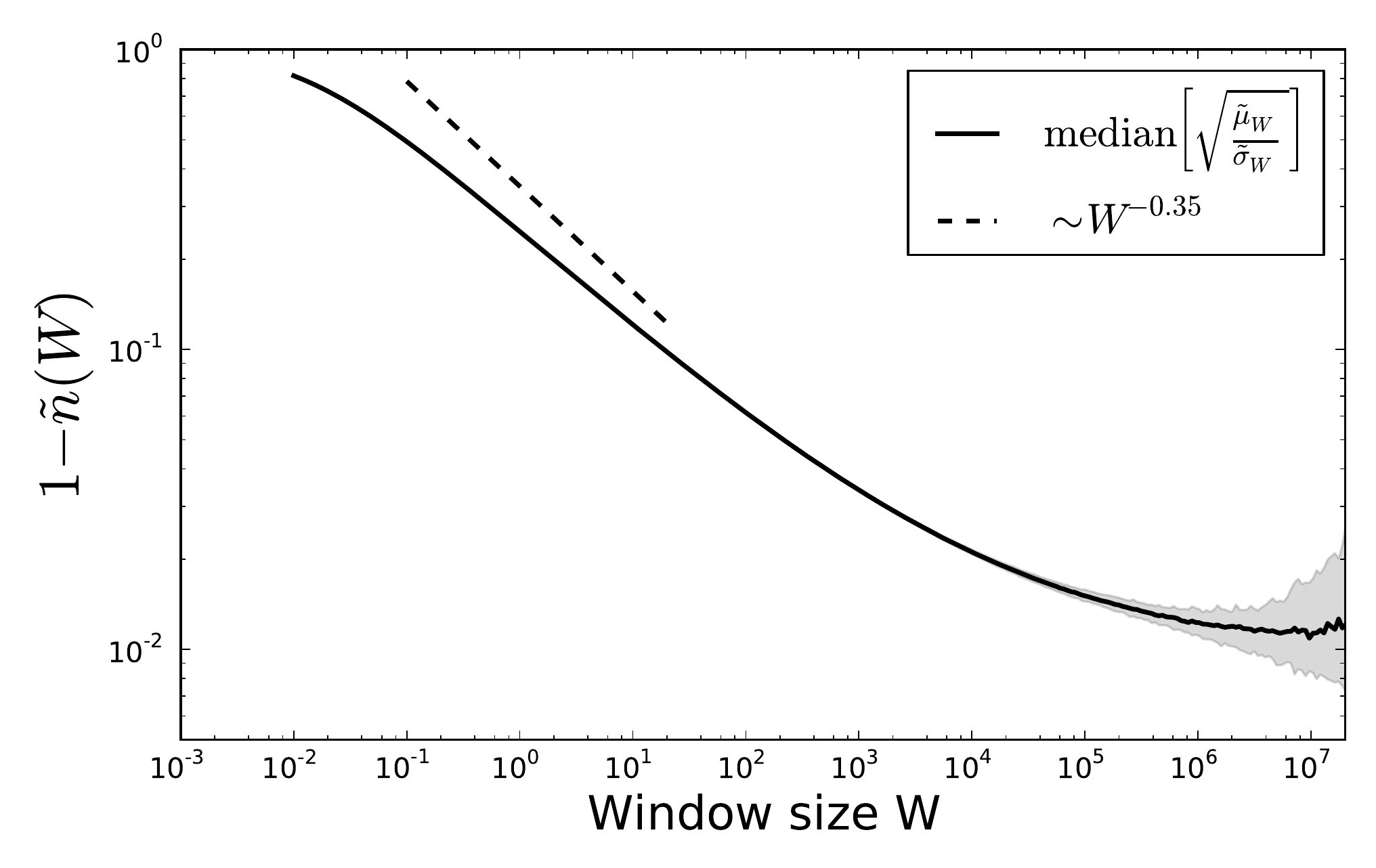}
\caption{\label{fig:power_law} $1 - \tilde{n}$ for simulated near-critical ($n = 0.99, \epsilon = 0.35$) power-law Hawkes processes. The value for $1-\tilde{n}$ that we obtain approximately scales as $\sim W^{-0.35}$ for large $W$. Note that the simulated process is only `near-critical' (with a branching ratio $n = 1 - 1 \times 10^{-2}$) so for very large $W$ the curve levels off and converges to $1 \times 10^{-2}$.}
\end{figure}

{
We have simulated such a process with exponent $\epsilon = 0.35$ and cut-off parameter $\tau_0 = 1.0$. To allow our simulation to attain a stationary state with an average event rate $\Lambda \approx 1$, we make our process very near critical by choosing $n = 0.99$ and $\mu = 0.01$. We simulate for a very long period $T = 1 \times 10^9$ and discard the first $0.9 \times 10^9$ to ensure the process is close to stationarity ($\Lambda > 0.99$ at $0.9 \times 10^9$). In Fig. \ref{fig:power_law} we present the results of branching ratio estimation using Eq. (\ref{eq:simple_estimator}) with a variety of window sizes $W$.
}

{
We note that the branching ratio estimate we obtain is very much dependent on the choice of window size used. To capture all the correlation present in the process and obtain estimates close to the the true input value $n = 0.99$ we must probe the correlation on very large scales, by choosing a very large value for $W$. The branching ratio obtained indeed varies with window size according to the law $W^{-\epsilon}$, in a similar way to the integral of the kernel
$\phi(\tau)$, see Fig. \ref{fig:power_law}.
}

\section{Empirical applications}
\subsection{Flash-crash revisited}

\begin{figure}[h]
\includegraphics[width=1\columnwidth]{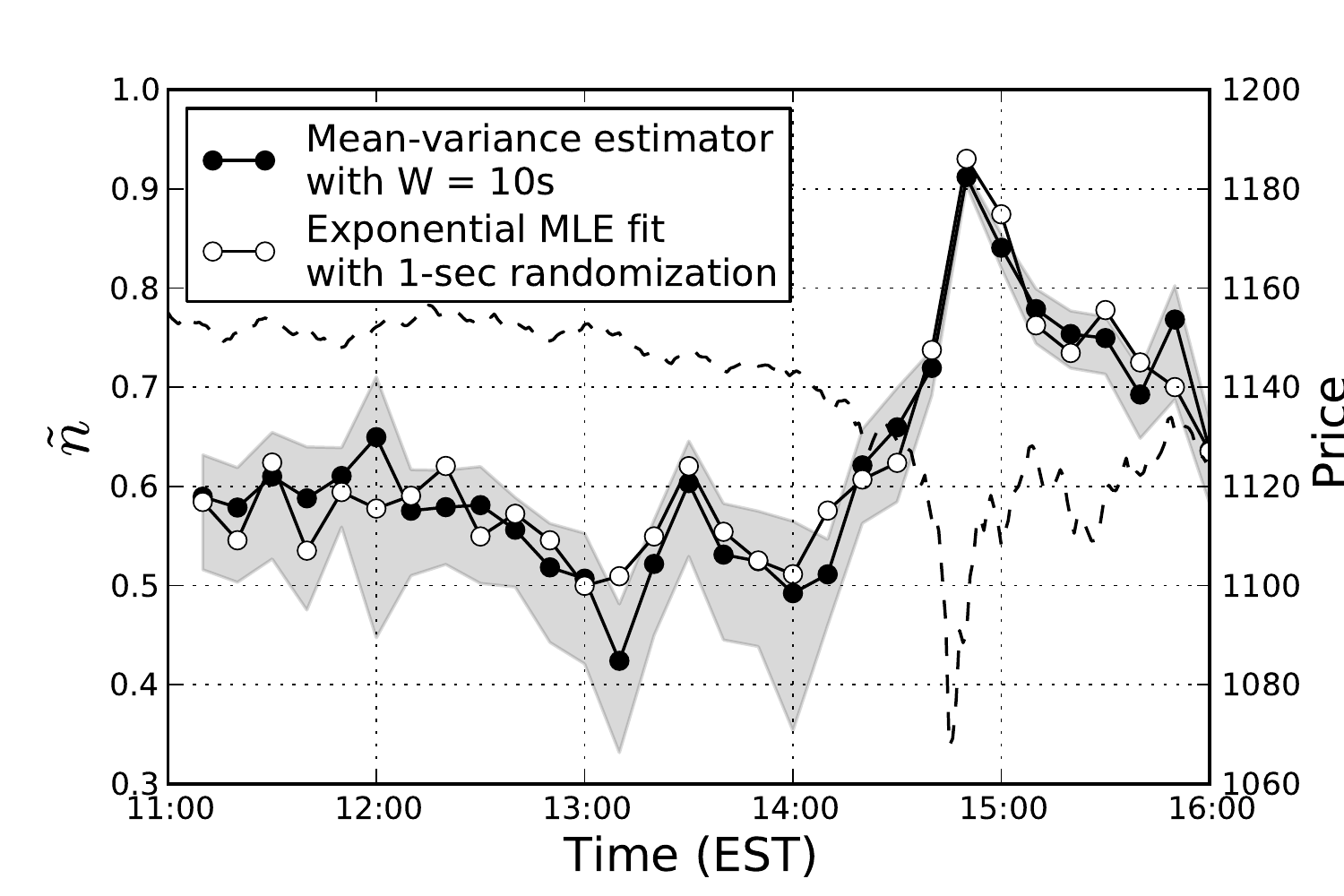}
\caption{\label{fig:flashcrash} Reproduction of the flash-crash branching ratio analysis of Filimonov \& Sornette using the mean-variance estimator. Our results compare well with those obtained by maximising the likelihood of the exponential model (with 1-second randomisation). The dashed line is the E-mini S\&P price. The points are placed at the end of the 10-minute period for which the branching ratio estimate is made. The shaded region is a [10\%,90\%] confidence interval generated by bootstrap re-sampling.}
\end{figure}

To demonstrate the effectiveness of this simple estimator we have repeated the flash-crash day branching ratio analysis of Filimonov \& Sornette \cite{filimonov}. We consider non-overlapping periods of 10 minutes in the hours of trading before and just after the flash-crash. For each 10-minute period, we calculate the sample mean and variance of the number of mid-price changes in the 60 windows of length $W=10s$ contained. The resulting values are plugged into Eq. (\ref{eq:simple_estimator}) to obtain an approximation of the branching ratio for each period. The results are presented in Fig. \ref{fig:flashcrash}.

Note that this simple estimator is biased, and for a general $W$, will {typically} underestimate the value of $\int_{-\infty}^\infty \nu(t) {\rm d}t$ and therefore the branching ratio. Since we consider a window size $W = 10s$ we have systematically underestimated $n$ in Figure \ref{fig:flashcrash} as measurements of $\nu(t)$ on this data have support outside the interval $[-10s,10s]$ --- there is still significant autocorrelation in the event rate at scales longer than $10s$.

However, we have identified that a window size $W$ of the order of approximately 10 to 30 seconds produces estimates of the branching ratio on our data in line with those obtained by ML fitting to the exponential model after intra-second randomisation (the method applied in \cite{filimonov}.) Note that we do not fix $\beta$ in our ML fit but let it settle naturally at the value which maximises the log-likelihood. We observe that this value $\hat{\beta}$ is very much dependent on the period of randomisation\footnote{To address the limited precision of the event data in \cite{filimonov}, the authors randomise timestamps uniformly inside the second that they are reported.} of the timestamps. When we randomise timestamps inside each millisecond we obtain $\hat{\beta}^{-1} \approx 10^{-2}$ for periods in 2010 but randomisation at larger intervals (e.g. the intra-second randomisation of Filimonov \& Sornette) prevents $\hat{\beta}^{-1}$ from exceeding values of the order of magnitude of the scale of randomisation.

\begin{figure}
\includegraphics[width=1\columnwidth]{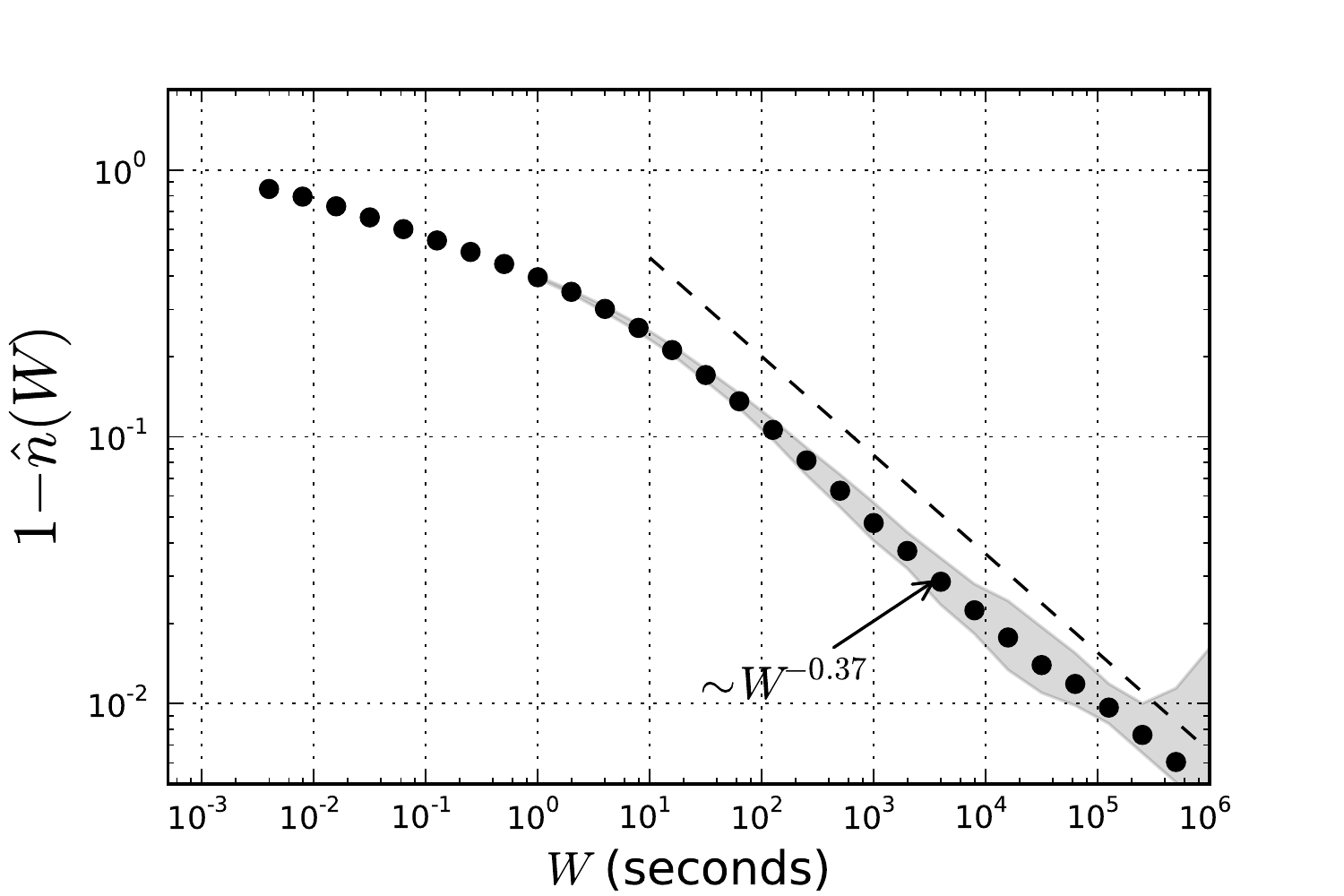}
\caption{\label{fig:n_2010} $1 - \tilde{n}$, the estimate of the branching ratio as a function of window size for E-mini S\&P mid-price changes in 2010. The mean and variance of $N_W$ are estimated on the full year. The change of power-law behaviour occurs around $100$ seconds. Note that we have `stitched' together all 5-minute bins of regular trading hours (09:30 to 16:00 EST) that contain at least one event (this solves a problem arising from missing data at half-days). We have then de-trended the intra-day seasonality by dividing each 5-minute event count by a normalised average event rate for each 5-minutes of the trading day calculated on the full year.}
\end{figure}

Note that since our results with $W = 10s$ correspond well with those obtained using the methods of Filimonov \& Sornette \cite{filimonov}, their procedure must \emph{also} underestimate the branching ratio. To converge on the true $n$ in expectation, we must take Eq. (\ref{eq:simple_estimator}) in the limit of $W \to \infty$. We do just this in Fig. \ref{fig:n_2010} for mid-price changes of the E-mini S\&P Futures contract in 2010. One notes that as the window size increases, the branching ratio converges to $n=1$ in a non-trivial way. As reported in \cite{criticalreflexivity} for the structure of the kernel at short and long time-scales, two regimes are detectable with a transition around five minutes. The branching ratio asymptotically tends towards $n=1.0$ with a scaling exponent $\epsilon = -0.37$ compatible with the value of $0.45$ estimated in \cite{criticalreflexivity} for a 14-year period. Note that in taking the limit $W \to \infty$ we consider variation in the event rate at significant time-scales to be explained by the stationary Hawkes model.

\subsection{Reflexivity : 1998 - 2013}

\begin{figure}
\includegraphics[width=1\columnwidth]{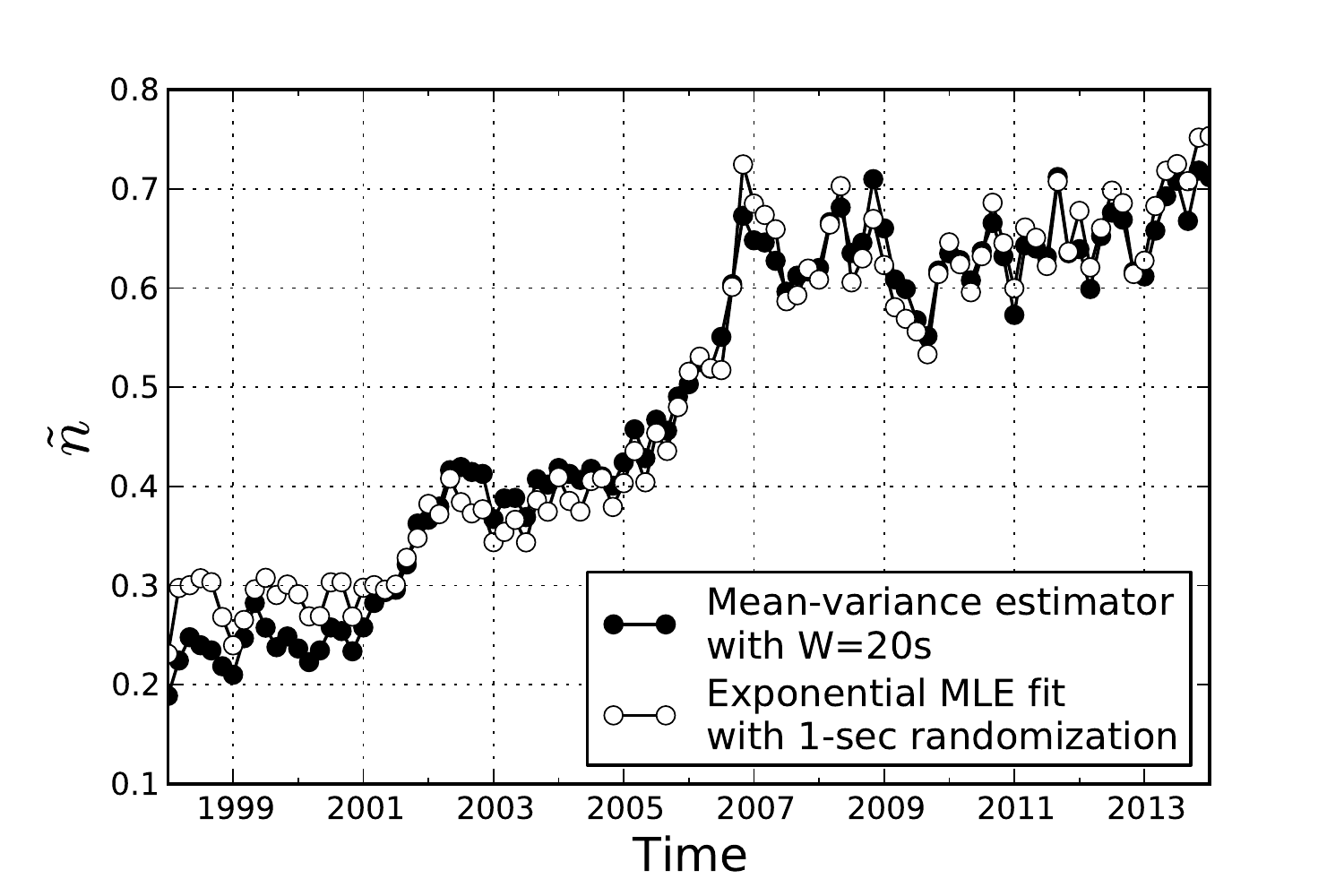}
\caption{\label{fig:19982013} Estimates of the branching ratio for 15-minute windows using the method of Filimonov \& Sornette \cite{filimonov} and estimates using the mean-variance estimator with a window size of $W=30s$. Note that our MLE results differ somewhat from the plot presented in \cite{filimonov}. We attribute this to differences in the data source or identification of mid-price changes.}
\end{figure}

Using the mean-variance estimator, we can also easily reproduce the result of \cite{filimonov} that claims to demonstrate that reflexivity has been increasing in the S\&P futures market since 1998. To do this we set our only parameter $W = 30s$ and estimate the branching ratio in periods of 15-minutes. In Fig. \ref{fig:19982013} we present the 2-month medians of these estimates beside the median of the branching ratio estimate obtained using the exponential maximum likelihood approach after intra-second randomisation.

\begin{figure}
\includegraphics[width=1\columnwidth]{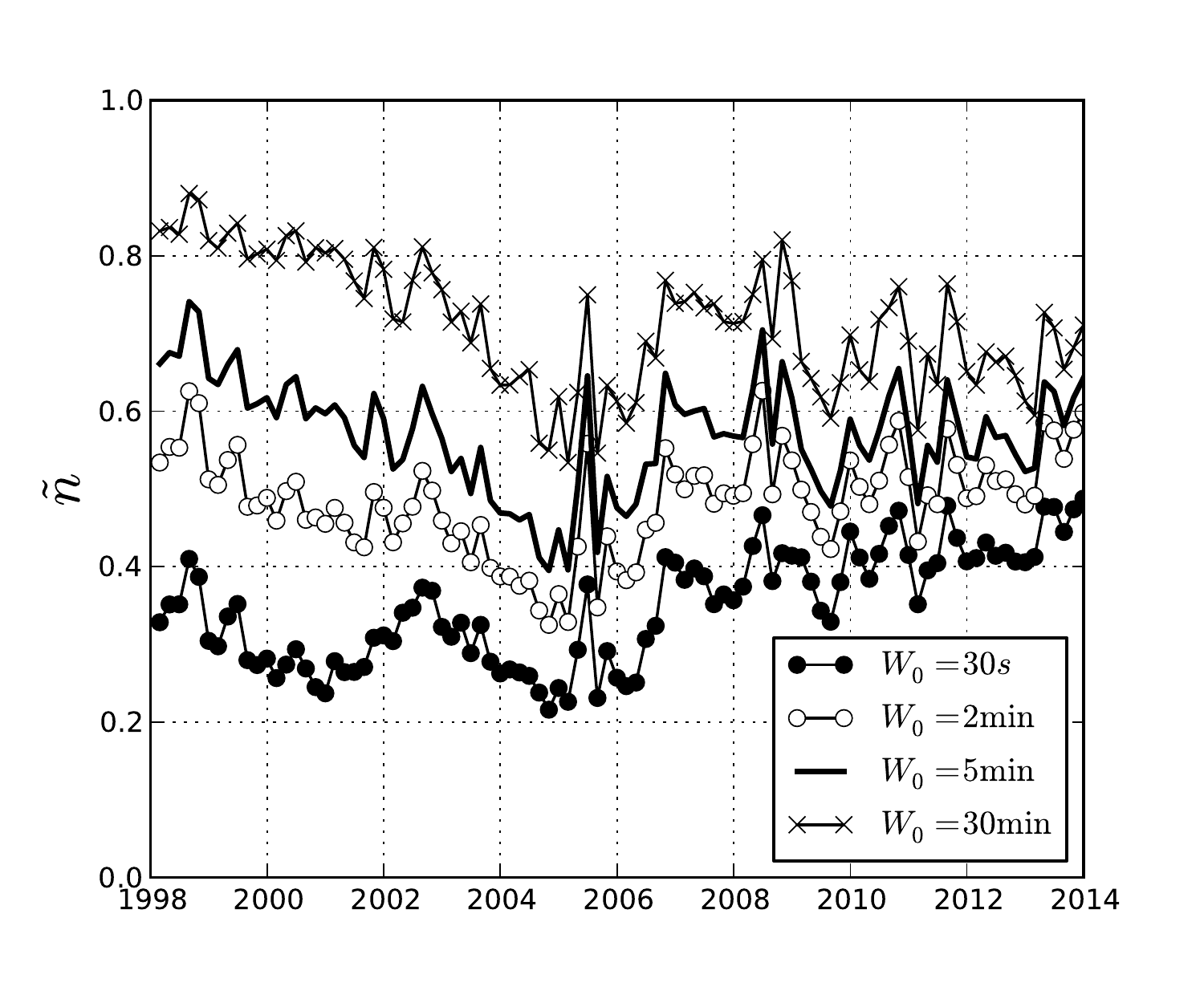}
\caption{\label{fig:varying_W} Estimates of the branching ratio on 2-month periods using the mean-variance estimator with a window size that follows Moore's law: $W_t = W_0 e^{-c(t-t_0)}$ with $c = -\log(1/2) / (18 \, \mathrm{ months})$ and $t_0=1998$. We again stitch together periods of regular trading hours and de-trend the event count by the intra-day U-shape for each year. When $W$ is appropriately rescaled, the branching ratio estimate is approximately constant through time, for all values of $W_0$, and tends to $n=1.0$ for large $W_0$.}
\end{figure}

Since we expect the minimum time-scale of correlation in the data to have decreased over the past decades (due to decreasing latency with advancing technology) we now re-perform the experiment but with a window size $W_t$ that follows Moore's law in such a way to keep the average number of events in $W_t$ roughly constant. More precisely the window size $W_t$ halves every 18 months; this describes well the increase in the high frequency activity of markets, see \cite{criticalreflexivity}. The results of this experiment are presented in Fig. \ref{fig:varying_W} and confirm that the kernel integral is approximately invariant over time provided that the measurement window $W_t$ is appropriately rescaled to account for the changing speed of interactions in the market. We find this quite remarkable, as this suggests that the amount of self-reflexivity in financial markets is {\it scale invariant}, and has not significantly 
increased due to high-frequency trading.

\section{Summary}

We have introduced a simple estimator for the branching ratio of Hawkes self-exciting point-process. The method is straight-forward to apply to empirical event data since it requires only a rudimentary calculation on the mean and variance of the event rate. The estimator does not suffer from the bias inherent to the contentious question of the choice of kernel in the likelihood maximisation approach, and furthermore it avoids the need for complicated numerical minimisation techniques.

Despite its simplicity, our estimator can accurately reproduce results obtained for the branching ratio using this prior method. The estimator presented is indeed a biased estimator, but it can be made asymptotically unbiased in the limit of large $W$ and $T$, for which we observe that the branching ratio for empirical mid-price changes of the E-mini S\&P futures contract approaches unity in line with previous theoretical and empirical results \cite{criticalreflexivity}.

Furthermore we demonstrate that if the window size is allowed to scale with Moore's law and adapt to the changing speed of interaction in the market over the past fifteen years, then the branching ratio approximation recovered is constant supporting prior observations of the invariance of the Hawkes kernel and branching ratio over time in \cite{criticalreflexivity}.

Finally, let us reiterate the caveat made above: the Hawkes analysis of the activity in financial markets leads to the conclusion that the process must be critical. However, it might well be that the dynamics of markets is more complicated and requires non-linearities absent from the Hawkes process defined by Eq. (\ref{eq:hawkes})). More work on this issue would be highly interesting, and is in progress in our group.

\section*{Acknowledgements} We thank V. Filimonov for many productive discussions about this topic. We are also indebted to N. Bercot, J. Kockelkoren, M. Potters,
I. Mastromatteo, P. Blanc and J. Donier for interesting and useful comments. 

\bibliography{branching}

\end{document}